\documentclass[aps,prd,preprintnumbers,showpacs,twocolumn,groupedaddress,nofootinbib]{revtex4}

\usepackage{graphicx}
\usepackage{latexsym}
\usepackage{xcolor}
\usepackage{ulem}

\def\beq{\begin{equation}}
\def\eeq{\end{equation}}
\def\bey{\begin{eqnarray}}
\def\eey{\end{eqnarray}}

\def\lsim{\mathrel{\raise.3ex\hbox{$<$\kern-.75em\lower1ex\hbox{$\sim$}}}}
\def\gsim{\mathrel{\raise.3ex\hbox{$>$\kern-.75em\lower1ex\hbox{$\sim$}}}}

\begin{document}

\title{Interpretations of the possible 42.7 GeV $\gamma$-ray line}
\author{Lei Feng$^{1,\ast}$, Yun-Feng Liang$^{1,2}$, Tie-Kuang Dong$^{1}$ and Yi-Zhong Fan$^{1,\ast}$}
\affiliation{$^1$Key Laboratory of Dark Matter and Space Astronomy, Purple Mountain Observatory, Chinese Academy of Sciences, Nanjing 210008, China}
\affiliation{$^2$University of Chinese Academy of Sciences, Beijing, 100012, China}


\begin{abstract}
Recently Liang et al. found a tentative line signal at about 43 GeV in the directions of sixteen nearby Galaxy Clusters. If arising from dark matter annihilation, the mass of the dark matter particles should be $m_{\chi}\sim 43$ GeV and the annihilation cross section $\langle \sigma v \rangle \sim 5\times10^{-28}(\overline{\rm BF}/10^{3})^{-1}~{\rm cm^3~s^{-1}}$ is needed, where $\overline{\rm BF}$ is the averaged boost factor of the annihilation signal of these Galaxy Clusters. In this work we discuss several models which could interpret these features and estimate the model parameters. Usually sizeable coupling parameter is needed. The $2\sigma$ upper limits on the cross section of the dark matter annihilation into various final states such as $b\bar{b}$, $\mu^+\mu^-$ and so on are also presented.
\end{abstract}

\pacs{95.35.+d; 95.85.Pw}
\maketitle

\section{Introduction}

Dark matter particles may annihilate with each other or alternatively decay and then produce high energy $\gamma$-rays and particle/antiparticle pairs such as electrons/positrons, protons/anti-protons, neutrinos/antineutrinos and so on \cite{Jungman1996,Bertone2005,Hooper2007,Feng2010,Fan2010}. Searching for these signals in the data of cosmic ray and gamma-ray is the goal of the dark matter indirect detection. In general, there are two kinds of $\gamma$-ray signals, one is the continual spectrum component and the other is the gamma-ray line(s).

As for the continual spectrum search, the main challenge is to reliably distinguish between the dark matter signal and astrophysical background, since some astrophysical processes can also produce continual GeV-TeV gamma-ray emission.
That is why the physical origin of Galactic GeV excess, though found in a set of independent approaches with high statistical significance \cite{Goodenough:2009gk,Vitale:2009hr,Hooper:2010mq,Hooper:2010im,Abazajian:2012pn,Gordon:2013vta,Huang:2013pda,Hooper:2013rwa,Daylan:2014rsa} and the presence of the signal is known to be not due to the improper modeling of the diffuse galactic gamma-ray emission \cite{zhou2015,weinger2015,huang2015,Ajello2015}, is in heavy debate.
The most widely examined interpretations of galactic center excess include the dark matter annihilation (such as \cite{Izaguirre,hooper2,bi2015,Alves}) and the astrophysical origin \cite{Abazajian:2012pn,2015arXiv150605124L,2015arXiv150605104B,yuan}.
This essential problem of backgrounds also motivates people to search for regions with very clean gamma-ray background, for instance, the spherical dwarf galaxies.
However, up to now the data analysis of possible gamma-ray emission from spherical dwarf galaxies only showed very weak emission from Reticulum 2 \cite{Geringer-Sameth2015} and Tucana III \cite{LiS2016}.

On the other hand, the GeV-TeV gamma-ray line, if reliably detected, will be the smoking-gun signature of dark matter particles since no known physical processes can produce the mono-energetic gamma-ray feature. The branching fraction of mono-energetic DM annihilation channels, however, is typically loop-suppressed and $\langle\sigma v\rangle_{\chi\chi\rightarrow \gamma\gamma} \sim (10^{-4}-10^{-1})\langle \sigma v \rangle$, where $\langle\sigma v\rangle_{\chi\chi\rightarrow \gamma\gamma}$ is the cross section for DM particle annihilation into a pair of $\gamma-$rays \cite{Bergstrom:1997fh,Cline:2012nw,Bertone:2009cb,Ullio:1997ke,bergstromkaplan,Choi:2012ap,Anchordoqui:2009bn,Mambrini:2009ad,
Dudas:2009uq,Chalons:2011ia,Jackson:2009kg,Profumo:2010kp,Feng2015,Gustafsson:2007pc,Bertone:2010fn,Acharya:2012dz,Goodman:2010qn,Lee:2012bq,
Dudas:2012pb,Kyae:2012vi}. Consequently the $\gamma-$ray line signal is expected to be weak.
Nevertheless, after the successful launch of the Fermi space gamma ray telescope (hereafter Fermi) \cite{Atwood2009}, scientists have been paying great attentions to the gamma-ray line search. In 2012, Bringmann {\it et al.}~\cite{Bringmann:2012vr} and Weniger~\cite{Weniger:2012tx} had found a tentative $\sim 130~{\rm GeV}$ gamma-ray line from the publicly available data of Fermi-LAT~\cite{data}. Su and Finkbeiner \cite{Su:2010qj} found that there may be two gamma ray lines at $\sim 114 ~{\rm GeV}$ and $\sim 130~{\rm GeV}$. To interpret this gamma ray line signature, the cross section of dark matter annihilating to $\gamma \gamma$ is about $\sim 10^{-27}~{\rm cm^3~s^{-1}}$. Buckley and Hooper \cite{hooper2012} pointed out that sizeable couplings and some kinds of resonance are needed to get the needed cross section. Also there are lots of models to interpret this gamma ray line signature such as charge scalar mediator model \cite{Cline:2012nw}, Semi-Annihilation dark matter model\cite{semi}, Two Component Dark Matter \cite{twocomponent} and so on. Hektor et al. \cite{Hektor2012} reported further though a bit weaker evidence for the $\sim 130$~GeV $\gamma-$ray line emission from galaxy clusters in Fermi-LAT data (see however \cite{Huang2012b,Anderson2016}). The latest Pass 8 Fermi-LAT data analysis however do not confirm the presence of $\sim 130$~GeV $\gamma-$ray line feature in either the Galactic center or Galaxy clusters \cite{Ackermann2013Line,Albert2015Line,LiS2016}.

Recently Liang et al. \cite{liang2016} analyzed the gamma-ray emission in the directions of 16 massive nearby Galaxy Clusters. Their main finding is a  tentative line signal at $\sim 42.7~{\rm GeV}$. The flux of gamma ray generated by dark matter annihilation can be described by
\begin{eqnarray}
\Phi(E) = \frac{1}{4\pi}\frac{\langle \sigma v \rangle_{\gamma \gamma}}{2m_{\chi}^2}2\delta(E-m_{\chi})J_{\rm f},
\end{eqnarray}
the J-factor is defined as
\begin{eqnarray}
J_{\rm f}=\int d\Omega \int_{l.o.s} dl \rho^2,
\end{eqnarray}
where $\rho$ represents the density distribution of dark matter particles. For the regular dark matter smooth distribution models (i.e., without introducing the so-called boost factor of dark matter annihilation due to the ``presence" of poorly-known sub-structures), the J-factors in the directions of the Galactic central regions are much larger than that of Galaxy Clusters. Therefore, the flux of mono-energetic gamma ray line in the directions of these Galaxy Clusters is not expected to be brighter than that from the Galactic center unless the boost factor (BF) of Galaxy Clusters are high up to ${\rm BF}\sim 10^{3}$. Indeed, with the latest Fermi-LAT Pass 8 data, even for an isothermal Galactic dark matter profile that yield the ``loosest" constraint on the gamma-ray line annihilation channel, a tight constraint is $\langle \sigma v \rangle_{\gamma \gamma}\leq 5\times 10^{-28}~{\rm cm^{3}~s^{-1}}$ at $m_\chi \sim 42.7$ GeV \cite{Albert2015Line,fermi2015iso}. Hence Liang et al. \cite{liang2016} speculated that $\langle\sigma v\rangle_{\gamma \gamma}\sim 5\times 10^{-28}~{\rm cm^{3}~s^{-1}}$ at $m_\chi \sim 42.7$ GeV and the averaged boost factor of the dark matter annihilation signal from these massive Galaxy Clusters is $\overline{{{\rm BF}}}\sim 10^{3}$. Such high BFs have been proposed in \citep{Gao2012} but in other literature just moderate BFs have been suggested for the Galaxy Clusters \cite{SP2014}. The main challenge of estimating ${\rm BF}$ is the still poorly-known relative abundance of subhalos and of their substructure properties. In principle, these information can be inferred from the N-body cosmological simulations. The problem is that currently the highest-resolution simulations fail to resolve the whole subhalo hierarchy and the estimates of BFs are based on extrapolations over several orders of magnitude \cite{SP2014}. Therefore, the actual values of BFs for Galaxy Clusters are still uncertain.

In this work, we adopt several models suggested in \cite{Bergstrom:1997fh,Cline:2012nw,hooper2012,semi,Ibarra2012} to interpret the tentative 42.7 GeV line and estimate the suitable parameters. The constraints on the dark matter annihilation cross section for various final states such as $b\bar{b}$, $\mu^+\mu^-$ and so on are also briefly discussed.

\section{Models}
As mentioned above, the cross section of dark matter annihilation that can reproduce the tentative gamma ray line observed in the directions of these Galaxy Clusters depends on the poorly known boost factors. In this work following \cite{liang2016} we simply assume an averaged value of the boost factors of the Galaxy Clusters $\overline{\rm BF}=1000$ and then have $\langle \sigma v \rangle_{\gamma \gamma}\sim 5\times10^{-28} ~{\rm cm^3~s^{-1}}$.  It should be noted that smaller cross section with larger booster factor would loose the constraints on the coupling parameter and mediator mass. However, the adopted $\overline{\rm BF}\sim 10^{3}$ is already quite optimistic \cite{Gao2012,SP2014} and it is less likely that $\langle \sigma v \rangle_{\gamma \gamma}$ can be much smaller than $5\times10^{-28} ~{\rm cm^3~s^{-1}}$.

\subsection{Model With Boson Mediator and Triangle Loop}

\begin{figure}
\includegraphics[width=80mm,angle=0]{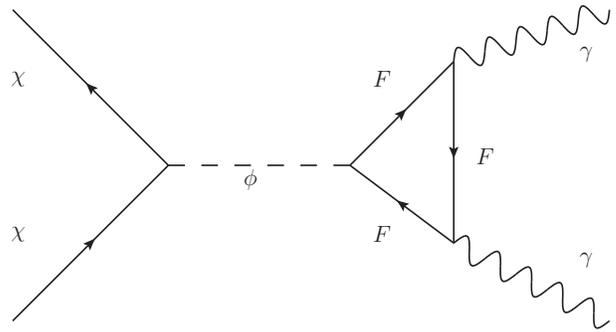}
  \caption{Dark matter annihilates into photons though scalar mediator and triangle fermion loop, where $\chi$ represents dark matter particles which are assumed to be Majorana fermions,  $\phi$ and $F$ denote the scalar mediator and the particle in the triangle loop, respectively.}
  \label{fig:1}
\end{figure}

The first model we consider is that dark matter particles annihilates to $\gamma \gamma$ through a boson propagator and triangle charged particle loop in $s$-channel~\cite{Bergstrom:1997fh}. The Feynman diagram is shown in Fig.~\ref{fig:1} (see also \cite{Bergstrom:1997fh,hooper2012,Basak,Duerr} for example). In this model, we assume that the dark matter particles are Majorana fermions. For this sort of diagrams, the low-velocity annihilation cross section is given by
\begin{eqnarray}
\langle \sigma v \rangle_{\gamma \gamma} &=& \frac{\alpha^2 g^2_{F} g^2_{\chi}}{256\pi^3}\frac{m^2_F}{[(4m^2_{\chi}-m^2_{\phi})^2+m^2_{\phi} \Gamma^2_{\phi}]} \nonumber \\
&\times& \bigg[\arctan[((m^2_F-m^2_\chi)/m^2_{\chi})^{-1/2}]  \bigg]^2,
\end{eqnarray}
where $\alpha=1/137$ is the fine-structure constant; $\Gamma_\phi$ and $m_\phi$ are the width and mass of the s-channel mediator $\phi$ respectively; $m_F$ and $m_\chi$ denote the mass of the charged particle $F$ in the triangle loop and dark matter particle $\chi$ respectively; $g_{\chi}$ and $g_{F}$ represent the couplings between the $s$-channel mediator and corresponding particles $\chi$ and $F$ respectively. Here we assume that $F$ carries unit charge. Usually the cross section is much smaller than the value of $5\times 10^{-28}~{\rm cm^3~s^{-1}}$, and hence below we focus on the resonance scenario with $m_{\chi}=1/2m_{\phi}=42.7~GeV$. Since the charged particles lighter than $\sim 100~{\rm GeV}$ have been strictly constrained by LEP II \cite{LEP}, we take  $m_F =100~{\rm GeV}$. Then we have
\begin{eqnarray}
\label{eq:2}
\langle\sigma v\rangle_{\gamma \gamma} \approx 5 \times 10^{-28} {\rm cm}^3/{\rm s} \, \bigg(\frac{g_{F} g_{\chi}}{1}\bigg)^2 \, \bigg(\frac{6.5 \, {\rm GeV}}{\Gamma_{{\phi}}}\bigg)^2.
\end{eqnarray}
This value is large enough to generate the observed 42.7 GeV gamma ray line (i.e., $\langle\sigma v\rangle_{\gamma \gamma} \sim 5\times 10^{-28}~{\rm cm^3~s^{-1}}$), with $\Gamma_\phi/g_{F} g_{\chi}=6.5~ {\rm GeV}$ and $m_F=100{\rm GeV}$.

\subsection{Model with Charged scalar as mediator}

\begin{figure}
\includegraphics[width=40mm,angle=0]{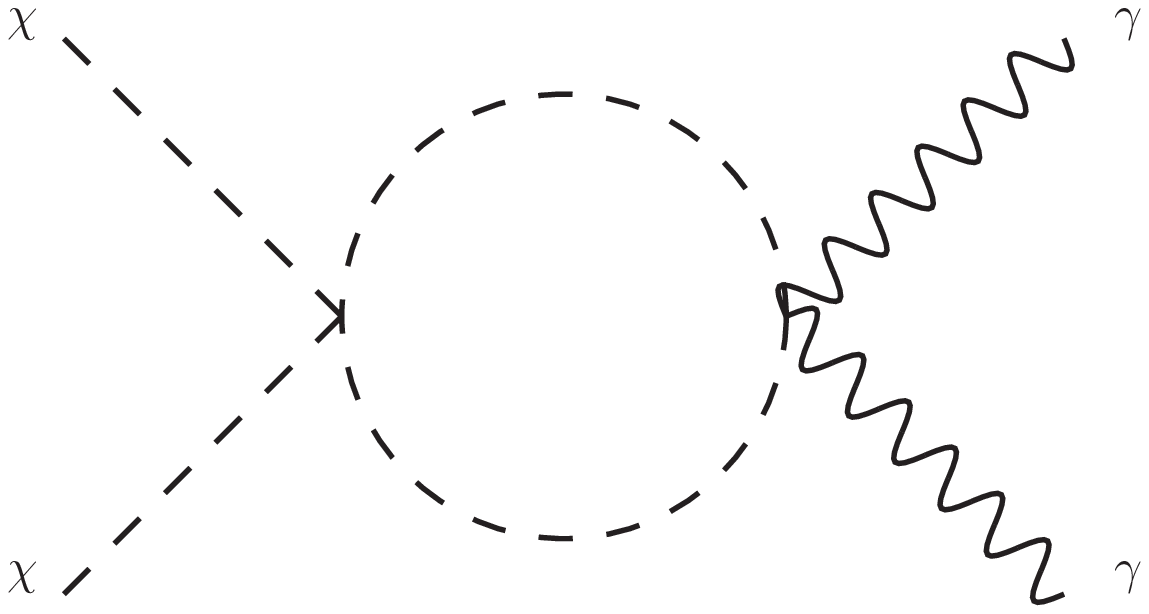}
\includegraphics[width=40mm,angle=0]{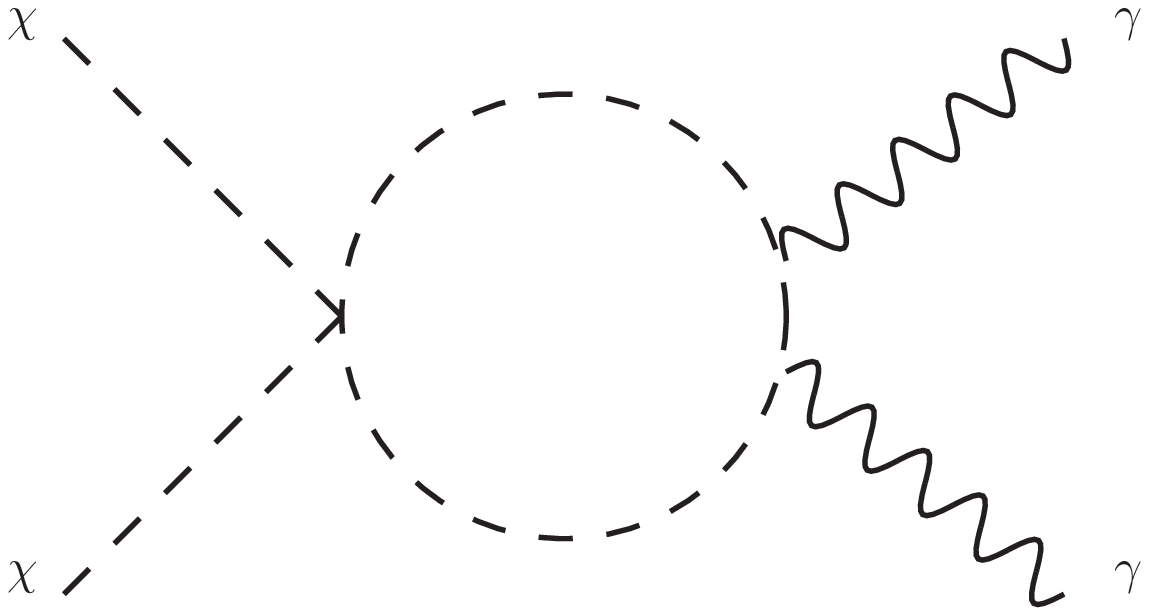}
  \caption{Scalar dark matter particles annihilate into two photons though charged scalar loop.}
  \label{fig:3}
\end{figure}

In this model, the scalar dark matter particles couple with charged scalar $S$ in the form of $\mathcal{L}=(\lambda_{\chi}/2) {\chi}^2 |S|^2$~\cite{Cline:2012nw} (see Fig.~\ref{fig:3}). Similarly to the previous subsection, we assume that $S$ carries unit charge. The low-velocity annihilation cross section is given by
\begin{eqnarray}
\langle\sigma v\rangle_{\gamma \gamma} = \frac{\alpha^2 \lambda^2_{\chi}}{128\pi^3 m^2_{\chi}}[1-(m^2_S/m^2_{\chi})\arcsin^2(m_{\chi}/m_S)]^2.
\end{eqnarray}
Similarly, we take the mass of extra charged particle to be 100 GeV (i.e., $m_{S}=100$ GeV), and then
\begin{eqnarray}
\langle\sigma v\rangle_{\gamma \gamma,{\rm max}} \approx 3.9 \times 10^{-31} \, {\rm cm}^3 ~{\rm s^{-1}} \bigg(\frac{\lambda_{\chi}}{1}\bigg)^2 \, \bigg(\frac{42.7\, {\rm GeV}}{m_{\chi}}\bigg)^2. \nonumber
\end{eqnarray}
Thus to get the required cross section with this model, the coupling parameter would be extremely large ($\lambda_{\chi} \sim 36$). Such a  coupling parameters $\lambda_\chi$ is so
large that the perturbative calculations are invalid and it will also induce a very low scale Landau pole.

\subsection{Semi-Annihilation dark matter model}

Finally, we consider the semi-annihilation dark matter model \cite{semi}. In this model, dark matter particles $V_i$ are dark gauge bosons for general dark gauge groups and dark matter particles annihilate into photons via the process $V_i V_j \rightarrow V_k \gamma$. If all types of dark matter particles have identical mass which labeled with $m_V$, the monochromatic photon energy is given by the following equation:
\begin{eqnarray}
E_{\gamma}=\frac{3}{4}m_{V}
\end{eqnarray}
To interpret this gamma ray line, the mass of dark matter particle is 56.9 GeV.

\begin{figure}
\includegraphics[width=55mm,angle=0]{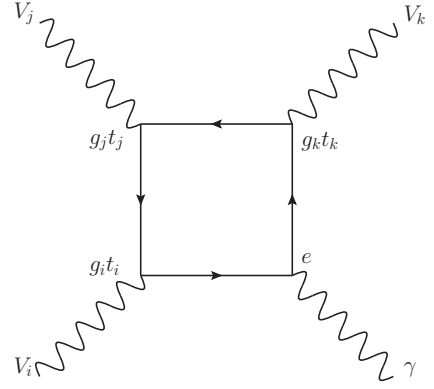}
  \caption{The diagram of photon production for dark matter semi-annihilation model.}
  \label{fig:4}
\end{figure}

The effective Lagrangian is described as \cite{semi}
\begin{eqnarray}
\mathcal{L} \supset &&  \sum_{24 \text{ perm}} \frac{g_i g_j g_k e}{180 (4 \pi)^2 M_M^4} \mathop{\rm Tr} [ t^i t^j t^k]  \\
&&\left(5 G^i_{\mu\nu} G^{j \nu\mu} G^k_{\lambda\rho} F^{\rho\lambda} - 14 G^i_{\mu\nu} G^{j \nu\lambda} G^k_{\lambda\rho} F^{\rho\mu} \right) ~~,\nonumber
\end{eqnarray}
where $M_M$ is the messenger's mass and the sum is over all possible twenty-four permutations of the four field strengths, $t^i$ are the generators of dark gauge group and $g_i$ are the gauge couplings. The Feynman diagram of dark matter semi-annihilation is shown in Fig.\ref{fig:4}.

In the dark SU(3) case, the semi-annihilation cross section is
\small
\begin{eqnarray}
\label{eq:sv}
&&\frac{1}{2} \langle{\sigma v}_s\rangle (V V\rightarrow V \gamma)  \nonumber \\
& =  & \frac{5}{192} \frac{1697}{460800\pi} \frac{ \alpha_d^3 \alpha}{M_V^2} N_f^2 \left( \frac{M_V}{M_M} \right)^8 F \left( \frac{M_M}{M_V} \right)  \label{eq:semisu3} \\ \nonumber
& \simeq & 5 \times 10^{-28} \, \text{cm}^3  / \text{s} \left( \frac{\alpha_d}{3.6} \right)^3 N_f^2 \left( \frac{100~GeV}{M_M} \right)^8 \left( \frac{M_V}{56.9~GeV} \right)^6, \nonumber
\end{eqnarray}
\small
where $\alpha_d$ is the coupling parameter, the function $F(M_V/M_M)$ is the form factor which includes the high order contribution of dark matter annihilation (more details can be found in \cite{semi}). Here we set $N_f=1$, meaning that there are three new charged fermions above the weak scale \cite{semi}. The cross section is divided by a factor of 2 because there is only one $\gamma-$ray has been generated per annihilation event. Here we also assume that the mass of extra charged particle is 100 GeV (i.e., $M_M=100$ GeV), with which a very large coupling ($\alpha_{d}\sim 3.6$) is needed to interpret the observed gamma ray line. Note that such a large coupling parameters is not perturbative and hence the approach is not self-consistent, implying that this model may be disfavored.

In this model, there is another gamma ray line at 56.9 GeV which is produced through dark matter annihilation into two photons. But the cross section of such a process is much smaller compare to dark matter semi-annihilation model\cite{semi}.

\subsection{Decaying intermediate particle model}
\begin{figure}
\includegraphics[width=55mm,angle=0]{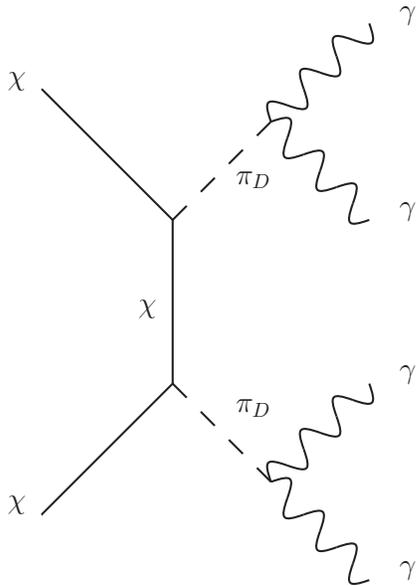}
  \caption{Photons production in decaying intermediate particle model.}
  \label{fig:decay}
\end{figure}

In this model, the dark matter particles $\chi$ first annihilate into neutral intermediate particles (dark pion for example, labeled with $\pi_D$ here) and then $\pi_D$ decays into two photons (as shown in Fig.~\ref{fig:decay}). The gamma-ray spectrum in this model is very broad when $\pi_D << m_\chi$. But when $m_\chi \sim m_{\pi_D}$, the produced intermediate particles would approximately at rest, leading to a line-like feature at $E_{\gamma} \sim m_\chi/2$~\cite{hooper2012,Ibarra2012,chu,bai,jjfan}. To generate the observed $\rm 42.7~GeV$ gamma ray line we need $m_\chi=\rm 85.4~GeV$.

The cross section for this scenario is given by:
\begin{eqnarray}
\sigma v_{\gamma \gamma} = \frac{g^4_{\chi}}{16 \pi m^2_\chi} \, \bigg[1-\bigg(\frac{m_{\pi_D}}{m_\chi}\bigg)^2\bigg]^{1/2}
\end{eqnarray}
where $m_\chi$ and $m_{\pi_D}$ are the mass of dark matter particle and the intermediate particle; $g_\chi$ represents the coupling between the dark matter and the dark pion. If we choose $m_{\pi_D}=\rm 82~GeV$, $g_\chi=\rm 0.087$ in order to get the cross section $\rm 5\times 10^{-28}~cm^3~s^{-1}$.

There is an extra gamma ray line at $\rm 85.4 GeV$ through s-channel annihilation via a
single virtual $\pi_D$ state. But the coupling between dark pion and photons is extremely small in many hidden sector models\cite{hooper2012}. Then the cross section of this process is usually very small and we do not consider it here.

\section{Discussions}
Motivated by the recent finding of a tentative gamma-ray at $\sim 42.7$ GeV in the direction of a group of nearby and massive Galaxy Clusters \cite{liang2016}, in this work we have examined the possible interpretations of such a signal. In total there are four models have been discussed. In the model with boson mediator ($\phi$) and triangle loop, a $<\sigma v>_{\gamma\gamma}\sim 5\times 10^{-28}~{\rm cm^{3}~s^{-1}}$ is possible supposing that $m_\phi=2m_\chi$ (i.e., the resonance), $\Gamma_\phi/g_{F} g_{\chi}=6.5~ {\rm GeV}$ and $m_F=100{\rm GeV}$. In the model with Charged scalar mediator, to yield $<\sigma v>_{\gamma\gamma}\sim 5\times 10^{-28}~{\rm cm^{3}~s^{-1}}$ a extremely large coupling constant $\lambda_{\chi} \sim 36$ is needed. In the semi-annihilation dark matter model, a sizable coupling constant $\alpha_d \sim 3.6$ is needed. In the decaying intermediate particle model, $g_\chi \sim \rm 0.087$ is needed to generate the needed cross section. Therefore, at least in principle the gamma-ray line signal at $\sim 42.7$ GeV seems possible. The extra charged particle ($F$ in II.A, $S$ in II.B and $M$ in II.C) can not be stable particles because there are extremely strong upper bounds on charged relics \cite{chargedrelics}. Here we consider the case that the new charged particle could quickly decay into Standard Model charged particles and (or) neutral massive particles(Z boson for example). The long-lived charge particles had been conclusively ruled out by the previously and currently collider experiments \cite{aad,chatrchyan,cms,cms2,atlas,abazov}. However, if
the mass splitting between charged and neutral particles is $\sim 0.1-1 {\rm GeV}$, it is extremely challenging to detect such particles at LHC \cite{perez,buckley}.

In view of that the global statistical significance of the $\sim 42.7$ GeV gamma-ray line signal is just $\sim 3\sigma$, further extensive tests of the reality of such a line will be crucially appreciated. Two ways are accessible in the near future. One is that Fermi-LAT continues to collect data from the Galaxy Clusters and the gamma-ray number in our regions of interest are expected to be doubled in 2020s. Then the possible statistical fluctuation origin of the $\sim 43$ GeV line can be directly tested. The other way is to use the gamma-ray data detected by new space missions with much better energy resolution. DArk Matter Particle Explorer (DAMPE \cite{dampe}) and  CALorimeteric Electron Telescope (CALET \cite{calet}), both launched in 2015 and are doing the sky survey,  provide us such opportunities. Though the effective areas of DAMPE and CALET are smaller than Fermi-LAT, the significantly improved energy resolution make them have prominent advantage in gamma-ray line search \cite{yeli}. It is expected that DAMPE and CALET can effectively test the possible statistical fluctuation origin or the instrumental origin of the $\sim 42.7$ GeV line signal.

Finally we would like to remark on the possible continuum spectrum component from the annihilation of dark matter particles yielding the gamma-ray line signal. In the stacking energy spectrum of gamma rays in the directions of the nearby 16 galaxy clusters \cite{liang2016}, single power law spectrum can well fit the emission below 40 GeV, hence we do not detect any significant ``continuum spectrum component from dark matter annihilation" and the dark matter annihilation cross section in different channels can be constrained. For such a purpose
we assume that the gamma-ray background can be modeled as a power-law with exponential cutoff function (PLE).
Then we use the model consisting of a PLE component, a gamma-ray line component, and an ``excess" continuum component to fit the stacking spectrum to derive maximal likelihood. Unbinned likelihood method similar to that adopted in \cite{liang2016} was used in the fitting, we refer readers to Sec.II of \cite{liang2016} for details of fitting procedure. For a series of $\langle{\sigma v}\rangle$ of continuum component (namely cross section of dark matter annihilating to leptons and quarks), we calculate the maximal likelihood $\mathcal{L}$ and find the cross section with $\ln{\cal L}=\ln{\cal L}_{\rm best}-1.35$ which is corresponding to $2\sigma$ upper limit. Here we got the $2\sigma$ upper limits of cross section of 42.7 GeV dark matter for various final states:
\begin{eqnarray}
\langle{\sigma v}\rangle_{b \bar{b}}~& < &~2.4\times10^{-23}~{\rm cm^3~s^{-1}}  / ~\overline{\rm BF} \nonumber \\
\langle{\sigma v}\rangle_{c \bar{c}}~& < &~2.6\times10^{-23}~{\rm cm^3~s^{-1}} /~ \overline{\rm BF}  \nonumber\\
\langle{\sigma v}\rangle_{\mu^+ \mu^{-}}~& < &~6.0\times10^{-23}~{\rm cm^3~s^{-1}} /~ \overline{\rm BF}  \nonumber \\
\langle{\sigma v}\rangle_{\tau^+ \tau^-}~& < &~2.3\times10^{-23}~{\rm cm^3~s^{-1}} /~ \overline{\rm BF} \nonumber
\end{eqnarray}
In calculations the NFW dark matter density profile \cite{nfw} is used. For $\overline{\rm BF}\sim 1000$, the $2\sigma$ upper limit of dark matter annihilating to $b\bar {b}$ ($c\bar {c}$) is about $<2.4\times10^{-26}~{\rm cm^3~s^{-1}}$ ($<2.6\times 10^{-26}~{\rm cm^3~s^{-1}}$). And for $\mu^+ \mu^{-}$ ($\tau^+ \tau^{-}$) channel, the $2\sigma$ upper limit is $<6.0 \times10^{-26}~{\rm cm^3~s^{-1}}$ ($<2.3\times10^{-26}~{\rm cm^3~s^{-1}}$). These constraints are comparable with the ones set by the dwarf galaxies \cite{fermi2015}.

\section*{Acknowledgments}

 This work was supported in part by 973 Programme of China under grant 2013CB837000, the National Natural Science of China under grants (Grant No. 11303096 and 11303107), the Youth Innovation Promotion Association CAS (Grant No. 2016288) and the Natural Science Foundation of Jiangsu Province (Grant No. BK20151608)

$^\ast$Email (fenglei@pmo.ac.cn, yzfan@pmo.ac.cn).

\end{document}